\documentclass[11pt, a4paper]{article}

\usepackage{jheppub}
\usepackage{amsmath, graphicx, braket, orcidlink, xcolor}
\usepackage{natbib}

\usepackage{lmodern,amsmath,multirow,cancel}
\usepackage{booktabs}
\usepackage{xstring}
\usepackage{ifthen}
\usepackage{ascmac,bm,mathrsfs,amsthm,amsfonts}
\usepackage{comment}
\usepackage[normalem]{ulem}

\begin{document}

\title{
Impact of dimension-8 SMEFT operators on baryogenesis via sphaleron decoupling
}

\author{Kiyoto Ogawa\,\orcidlink{0009-0001-7115-7107}\,$^{a}$ and Masanori Tanaka\,\orcidlink{0000-0002-1303-7043}\,$^{b}$}
\emailAdd{ogawa.kiyoto.f8@s.mail.nagoya-u.ac.jp}
\affiliation{$^{a}$Department of Physics, Nagoya University, Furo-cho Chikusa-ku, Nagoya, 464-8602, Japan}

\emailAdd{tanaka@pku.edu.cn}
\affiliation{$^{b}$Center for High Energy Physics, Peking University, Beijing 100871, China}


\abstract{
We investigate whether a baryogenesis mechanism known as sphalerogenesis can account for the observed baryon asymmetry of the Universe within the Standard Model effective field theory.
In this scenario, the baryon asymmetry is generated through the $CP$-asymmetric decoupling of electroweak (EW) sphaleron-like transitions.
We introduce seven $CP$-violating dimension-8 operators constructed from the Higgs doublet and the $SU(2)_L$ gauge fields and show that five of them can individually account for the observed baryon asymmetry with satisfying experimental constraints from colliders or electron electric dipole moment measurements.
We further study their impact in the presence of a $CP$-violating dimension-6 operator.
We find that the dimension-8 contributions can be comparable to the dimension-6 contribution when the dimension-8 operators are generated at one loop, demonstrating that loop-order counting can be as important as canonical mass-dimension counting in sphalerogenesis.
}

\maketitle
\flushbottom


\section{Introduction}

One of the major achievements of standard cosmology is the successful explanation of the primordial abundances of light elements through big-bang nucleosynthesis (BBN).
In the standard BBN framework, these abundances depend sensitively on the baryon-to-entropy ratio $n_{B}/s$.
Precise measurements of the angular power spectrum of the cosmic microwave background (CMB) constrain this ratio to be~\cite{Planck:2018vyg, ParticleDataGroup:2024cfk}
\begin{align}
\label{eq:BAU}
8.41 \times 10^{-11}
<
\frac{n_{B}}{s}
<
8.75 \times 10^{-11} \,.
\end{align}
Measurements of the primordial deuterium abundance in high-redshift quasar absorption systems provide a consistent constraint on $n_{B}/s$~\cite{Yeh:2022heq}.
Explaining this observed baryon asymmetry of the Universe (BAU) requires a mechanism that dynamically generates a nonzero baryon number.
Such a mechanism must satisfy the three Sakharov conditions: (i) baryon-number violation, (ii) $C$ and $CP$ violation, and (iii) departure from thermal equilibrium~\cite{Sakharov:1967dj}.

Electroweak baryogenesis within the Standard Model (SM) was originally proposed as a possible explanation of the BAU~\cite{Kuzmin:1985mm}.
However, lattice simulations have shown that, for the observed Higgs boson mass, the electroweak (EW) phase transition in the SM is a smooth crossover rather than a first-order phase transition~\cite{Kajantie:1996mn,Csikor:1998eu,Aoki:1999fi,DOnofrio:2014rug,DOnofrio:2015gop}.
Consequently, the departure from thermal equilibrium required in conventional EW baryogenesis cannot be realized within the SM.
Moreover, the amount of $CP$ violation contained in the Cabibbo--Kobayashi--Maskawa (CKM) matrix is insufficient to generate the observed baryon asymmetry in Eq.~\eqref{eq:BAU}~\cite{Gavela:1994dt,Huet:1994jb}.
These difficulties strongly motivate extensions of the SM.

Recently, an alternative baryogenesis mechanism called sphalerogenesis has been proposed~\cite{Kharzeev:2019rsy,Hong:2023zrf,Tanaka:2025cpw,Ogawa:2026olw}.
In this scenario, the BAU is generated through the $CP$-asymmetric decoupling of EW sphaleron-like transitions during the smooth EW symmetry breaking.
The feasibility of sphalerogenesis has been investigated within the framework of the Standard Model effective field theory (SMEFT)~\cite{Tanaka:2025cpw}.
When we consider operators up to dimension-6, only the EW-Weinberg operator defined in Eq.~\eqref{eq:Scp6} provides the $CP$-violating source required for sphalerogenesis.
Motivated by this observation, it was subsequently demonstrated that extensions of the SM containing $SU(2)_{L}$ multiplet fields can provide plausible ultraviolet (UV) completions capable of reproducing the observed baryon asymmetry through sphalerogenesis~\cite{Ogawa:2026olw}.

Previous studies of sphalerogenesis have focused primarily on the EW-Weinberg operator~\cite{Tanaka:2025cpw,Ogawa:2026olw}.
However, restricting the analysis to dimension-6 operators is not necessarily sufficient for a consistent SMEFT treatment.
In the context of conventional EW baryogenesis, dimension-8 operators have been shown to modify the predicted baryon asymmetry significantly and therefore can be essential for assessing the validity of the EFT expansion~\cite{deVries:2017ncy}.
It is thus important to investigate whether higher-dimensional operators beyond the EW-Weinberg operator can also generate the observed BAU through sphalerogenesis.

Baryon asymmetry generation near the EW scale has also been studied within the SMEFT framework in several other contexts, including conventional EW baryogenesis~\cite{Bodeker:2004ws,Huber:2006ri,Huang:2015izx,deVries:2017ncy} and baryogenesis through the chiral magnetic effect~\cite{Liu:2024mdo}.
These scenarios rely on a first-order EW phase transition, whereas sphalerogenesis occurs during the smooth EW crossover.
Therefore, sphalerogenesis provides a complementary framework for EW-scale baryogenesis.
For the reasons outlined above, a systematic identification of the higher-dimensional operators capable of generating the observed BAU through sphalerogenesis is important.

In this paper, we investigate sphalerogenesis in the SMEFT including $CP$-violating dimension-8 operators.
In Refs.~\cite{Nauta:2000xi,Nauta:2002ru}, it was shown that two dimension-8 operators can induce $CP$ violation in sphaleron transitions.
In addition to these two operators, we systematically analyze five additional dimension-8 operators constructed from the SM Higgs doublet and the $SU(2)_{L}$ gauge fields~\cite{Remmen:2019cyz,Li:2020gnx,Kondo:2022wcw,Naskar:2022rpg,Corbett:2024yoy,Murphy:2020rsh}.
We show that three of these additional operators also contribute to sphalerogenesis, so that five of the seven operators considered can generate the observed baryon asymmetry.
We further quantify the impact of the dimension-8 operators in the presence of the EW-Weinberg operator.
In particular, we show that dimension-8 operators generated at one loop can contribute at a level comparable to the two-loop-induced EW-Weinberg operator, demonstrating the importance of loop-order counting in addition to canonical mass-dimension counting.

The remainder of this paper is organized as follows.
In Section~\ref{sec:setup}, we introduce the SMEFT operators considered in our analysis.
In Section~\ref{sec:Sph_CP}, we define the EW sphaleron field configuration and its reduced action, and derive the contributions of the dimension-8 operators to the $CP$ asymmetry in sphaleron-like transitions.
In Section~\ref{sec:BoltzmannEq}, we formulate the Boltzmann equation for baryon asymmetry and present the numerical results for each dimension-8 operator.
In Section~\ref{sec:UVcompletion}, we discuss the implications of our results for UV matching, with particular emphasis on the relative importance of dimension-6 and dimension-8 contributions.
Further phenomenological and theoretical considerations are presented in Section~\ref{sec:discussions}.
Section~\ref{sec:conclusions} summarizes our conclusions.


\section{Setup}
\label{sec:setup}

We describe the SMEFT framework adopted in this work.
Throughout our analysis, we use the metric convention $\eta_{\mu\nu} = \operatorname{diag}(-1,1,1,1)$.
The action is defined by
\begin{align}
\label{eq:tot_action}
S = S_{\rm SM} + S_{CP}^{(6)} + S_{CP}^{(8)} \,.
\end{align}
The SM contribution is given by
\begin{align}
\label{eq:Ssm}
S_{\rm SM} = \int d^4x
\left[
-\frac{1}{2} \operatorname{tr}
\left( W_{\mu\nu}W^{\mu\nu} \right) - (D_\mu\Phi)^\dagger D^\mu \Phi - V_{0}(\Phi)
\right] \,,
\end{align}
where $W_{\mu\nu}$ denotes the $SU(2)_L$ gauge field strength and $\Phi$ is the SM Higgs doublet.
The Higgs potential is defined as
\begin{align}
V_{0}(\Phi) = -\mu^2\Phi^\dagger\Phi + \lambda(\Phi^\dagger\Phi)^2 \,.
\end{align}
We note that the hypercharge gauge field contribution is neglected in our analysis because its effect on the sphaleron configuration is known to be small~\cite{Klinkhamer:1984di,Klinkhamer:1990fi}.

The second term in Eq.~\eqref{eq:tot_action} contains the $CP$-violating dimension-6 contribution.
Within the bosonic operator sector considered here, it was shown in Ref.~\cite{Tanaka:2025cpw} that the only dimension-6 operator that contributes to sphalerogenesis is the EW-Weinberg operator~\cite{Banno:2024apv}
\begin{align}
\label{eq:Scp6}
S_{CP}^{(6)}
=
- \int d^4x \frac{g}{3\Lambda_{\rm dim6}^2} \epsilon^{abc} \widetilde W_{\mu\nu}^{a} W^{b\nu\rho} W^{c\mu}_{\rho} \,,
\end{align}
where $\widetilde W^{a\mu\nu} = \frac{1}{2} \epsilon^{\mu\nu\rho\sigma} W_{\rho\sigma}^{a}$ with $\epsilon^{0123}=+1$. 
Here, Greek indices denote Lorentz indices, while Latin indices $a,b,c=1,2,3$ denote $SU(2)_L$ adjoint indices.
The quantities $g$ and $\epsilon^{abc}$ are the $SU(2)_L$ gauge coupling and structure constants, respectively.

The dimension-8 contribution in Eq.~\eqref{eq:tot_action} is defined as
\begin{align}
\label{eq:Action_dim8}
S_{CP}^{(8)}
= \int d^4x \sum_{i=1}^{7} \frac{c_i}{\Lambda_{\rm dim8}^4} \mathcal O_i \,,
\end{align}
where $\mathcal O_i$ and $c_i$ denote the $CP$-violating dimension-8 operators and their corresponding Wilson coefficients, respectively~\cite{Remmen:2019cyz,Li:2020gnx,Kondo:2022wcw,Naskar:2022rpg,Corbett:2024yoy,Murphy:2020rsh}.
We restrict our analysis to the following $CP$-odd bosonic operators constructed from the Higgs doublet and the $SU(2)_L$ gauge-field strength:
\begin{align}
\label{eq:dim8_OPEs}
\begin{aligned}
\mathcal O_1
&=
W_{\mu\nu}^{a} \widetilde W^{a\mu\nu} (D^\rho\Phi)^\dagger D_\rho\Phi \,,
\\
\mathcal O_2
&=
i\epsilon^{abc} W^{a[\mu}{}_{\lambda} \widetilde W^{b\nu]\lambda} (D_\mu\Phi)^\dagger \tau^c D_\nu\Phi \,,
\\
\mathcal O_3
&=
\epsilon^{abc} W_{\mu\nu}^{a} W^{b\mu}{}_{\lambda} \widetilde W^{c\nu\lambda} \Phi^\dagger\Phi \,,
\\
\mathcal O_4
&=
W^{a\mu\nu}W_{\mu\nu}^{a} W^{b\rho\sigma}\widetilde W_{\rho\sigma}^{b} \,, \\
\mathcal O_5
&=
W^{a\mu\nu}W_{\mu\nu}^{b}W_{\rho\sigma}^{a}\widetilde W^{b\rho\sigma} \,,
\\
\mathcal O_6
&=
i\widetilde W^{a\mu\nu} (\Phi^\dagger\Phi) (D_\mu\Phi)^\dagger \tau^a D_\nu\Phi \,,
\\
\mathcal O_7
&=
i\widetilde W^{a\mu\nu} (\Phi^\dagger\tau^a\Phi) (D_\mu\Phi)^\dagger D_\nu\Phi \,,
\end{aligned}
\end{align}
The antisymmetrization convention used in Eq.~\eqref{eq:dim8_OPEs} is
\begin{align}
W^{a[\mu}{}_{\lambda}
\widetilde W^{b\nu]\lambda}
\equiv
W^{a\mu}{}_{\lambda}
\widetilde W^{b\nu\lambda}
-
W^{a\nu}{}_{\lambda}
\widetilde W^{b\mu\lambda}.
\end{align}
The matrices $\tau^a$ $(a=1,2,3)$ are the Pauli matrices.


\section{Dynamics of the $CP$-violating EW sphaleron decoupling}
\label{sec:Sph_CP}

\subsection{Sphaleron ansatz}

We begin by reviewing the non-contractible-loop ansatz used to describe the EW sphaleron configuration.
In the SM, a field configuration defined on the two-sphere at spatial infinity $(S^2)$ can be continuously deformed to the vacuum configuration because the corresponding mapping to the Higgs vacuum manifold $(S^3)$ is topologically trivial.
Nevertheless, a nontrivial topology emerges when one considers a non-contractible loop in the field-configuration space that connects topologically distinct adjacent vacua~\cite{Manton:1983nd}.
The loop parameter, together with the spatial two-sphere, defines the mapping $S^{2+1}\to S^3$.

The corresponding ansatz for the gauge and Higgs fields is given by~\cite{Manton:1983nd}
\begin{align}
\label{eq:sph_ansatz}
\begin{aligned}
&W_{\mu}(\mu,r,\theta,\phi)\,dx^{\mu} = -\frac{i}{g}f(r)\,dU_{\infty}U_{\infty}^{-1} \,,
\\
&\Phi(\mu,r,\theta,\phi)
=
\frac{v}{\sqrt{2}} \left[1-h(r)\right] 
\begin{pmatrix}
0\\
e^{-i\mu}c_{\mu}
\end{pmatrix}
+
\frac{v}{\sqrt{2}} h(r)U_{\infty}
\begin{pmatrix}
0\\
1
\end{pmatrix} \,,
\end{aligned}
\end{align}
where $c_x\equiv\cos x$ and $s_x\equiv\sin x$.
The variables $(r,\theta,\phi)$ denote the spatial coordinates, while $\mu$ parametrizes the non-contractible loop.
The neighboring vacuum configurations are located at the endpoints of the loop, whereas the sphaleron configuration corresponds to the saddle point at $\mu=\pi/2$.

The profile functions $f(r)$ and $h(r)$ describe the radial dependence of the $SU(2)_L$ gauge and Higgs fields, respectively, and satisfy the following boundary conditions
\begin{align}
f(0)=h(0)=0 \,, 
\qquad
\lim_{r\to\infty}f(r) = \lim_{r\to\infty}h(r) = 1 \,.
\end{align}
Their precise forms can be obtained by numerically solving the equations of motion for the gauge and Higgs fields (for example, see Refs.~\cite{Klinkhamer:1984di,Gan:2017mcv,Kanemura:2020yyr}).
Instead of using the full numerical solutions, we adopt the following approximate profile functions~\cite{Klinkhamer:1984di}:
\begin{align}
\label{eq:fb_hb}
f(\xi)
&=
\begin{cases}
\displaystyle
\frac{\xi^2}{\Xi(\Xi+4)},
& \xi\leq\Xi,
\\[2mm]
\displaystyle
1-\frac{4}{\Xi+4}
\exp\left(\frac{\Xi-\xi}{2}\right),
& \xi>\Xi,
\end{cases}
\\
h(\xi)
&=
\begin{cases}
\displaystyle
\frac{\sigma\Omega+1}{\sigma\Omega+2}
\frac{\xi}{\Omega},
& \xi\leq\Omega,
\\[2mm]
\displaystyle
1-\frac{\Omega}{\sigma\Omega+2}
\frac{1}{\xi}
\exp\left[\sigma(\Omega-\xi)\right],
& \xi>\Omega,
\end{cases}
\end{align}
where $\sigma=\sqrt{2\lambda/g^2}$ and $\xi=gvr$. 
The parameters $\Xi$ and $\Omega$ are determined by minimizing the bosonic energy functional at $\mu=\pi/2$, corresponding to the saddle point between topologically distinct adjacent vacua~\cite{Manton:1983nd,Klinkhamer:1984di}.
We numerically find
\begin{align}
\label{eq:Xi0_Omega0}
\Xi=1.467\equiv\Xi_0 \,, \quad \Omega=1.701\equiv\Omega_0 \,.
\end{align}
In the following, we refer to the configuration specified by Eq.~\eqref{eq:Xi0_Omega0} as the true sphaleron configuration.
The $2\times2$ matrix $U_{\infty}$ appearing in Eq.~\eqref{eq:sph_ansatz} is given by~\cite{Manton:1983nd}
\begin{align}
U_{\infty}
=
\begin{pmatrix}
e^{i\mu}(c_{\mu}-is_{\mu}c_{\theta}) & e^{i\phi}s_{\mu}s_{\theta}
\\
-e^{-i\phi}s_{\mu}s_{\theta} & e^{-i\mu}(c_{\mu}+is_{\mu}c_{\theta})
\end{pmatrix} \,.
\end{align}

In the finite-temperature analysis, the zero-temperature Higgs vacuum expectation value $v$ is replaced by the temperature-dependent value $v(T)$.
We use the fitting formula obtained from lattice simulations~\cite{DOnofrio:2014rug,Kharzeev:2019rsy}
\begin{align}
v(T) \simeq 3T\sqrt{1-\frac{T}{T_{\rm EW}}} \,,
\end{align}
where the SM crossover temperature is taken to be $T_{\rm EW}\simeq159.5~{\rm GeV}$~\cite{DOnofrio:2014rug,DOnofrio:2015gop,Annala:2023jvr}.
At finite temperatures, the dimensionless radial coordinate is taken as $\xi=g v(T)r$. 

\subsection{Reduced sphaleron action}

We next introduce the reduced action describing field configuration evolution along the non-contractible loop.
This action is used to determine the $CP$ asymmetry entering the Boltzmann equation for the baryon number.
We here employ the reduced description of sphaleron transitions developed in Refs.~\cite{Aoyama:1987nd,Funakubo:1991hm,Funakubo:1992nq,Nauta:2000xi,Nauta:2002ru,Tye:2015tva,Tye:2016pxi,Tye:2017hfv,Qiu:2018wfb}.

Substituting the ansatz in Eq.~\eqref{eq:sph_ansatz} into the action in Eq.~\eqref{eq:tot_action}, we obtain
\begin{align}
\label{eq:Ssph}
S_{\rm sph}
=
\int dt
\left[
\frac{M(\mu, T)}{2}
\left(\frac{d\mu}{d\eta}\right)^2 + G(\mu, T) \left(\frac{d\mu}{d\eta}\right)^3 - V(\mu, T) \right] \,,
\end{align}
where $\eta=g v(T)t$. 
The explicit expressions for $M(\mu, T)$ and $V(\mu, T)$ are given in the appendix of Ref.~\cite{Ogawa:2026olw}.
The terms with linear $d\mu/d\eta$ are neglected in our analysis because those can be expressed as a total time derivative and therefore do not affect the sphaleron transition rate. 
The term proportional to $G(\mu)$ is odd under reversal of the direction of motion along the non-contractible loop.
A nonvanishing $G(\mu)$ therefore induces different transition probabilities for motion in the positive and negative $\mu$ directions~\cite{Nauta:2000xi,Nauta:2002ru}.
This difference represents a $CP$ asymmetry between sphaleron transitions toward neighboring vacua.

Following the formalism of Ref.~\cite{Ogawa:2026olw}, we quantify the effective $CP$ asymmetry as
\begin{align}
\label{eq:Acp_eff}
A_{CP}^{\rm eff}(T)
=
\frac{3\pi}{4}\kappa_{CP}
\sqrt{\frac{8T}{\pi}}
\frac{
G(\mu=\pi/2, T)
}{
M^{3/2}(\mu=\pi/2, T)
}.
\end{align}
Here, $\kappa_{CP}$ parametrizes the uncertainty associated with the reduced description.
We adopt $1 \leq \kappa_{CP} \leq 4$ as representative benchmark values.

The SM action alone gives $G(\mu, T)=0$. 
By contrast, the $CP$-violating contributions $S_{CP}^{(6)}$ and $S_{CP}^{(8)}$ can generate a nonzero cubic term.
For the EW-Weinberg operator, we obtain~\cite{Tanaka:2025cpw,Ogawa:2026olw}
\begin{align}
\label{eq:G_dim6}
G_{\rm dim6}(\mu, T)
=
\frac{256\pi}{45}
g v(T)
s_{\mu}^2
\left(4-s_{\mu}^2\right)
\left(
\frac{v(T)}{\Lambda_{\rm dim6}}
\right)^2.
\end{align}

The dimension-8 contribution to the reduced action takes the following form
\begin{align}
\label{eq:Ssph_dim8}
S_{CP}^{(8)}
=
\sum_{i=1}^{7}
c_i
\int dt\,
G_i(\mu, T)
\left(\frac{d\mu}{d\eta}\right)^3.
\end{align}
Substituting Eqs.~\eqref{eq:sph_ansatz} and \eqref{eq:Action_dim8} into the action, we find that $\mathcal O_1$, $\mathcal O_2$, $\mathcal O_3$, $\mathcal O_4$, and $\mathcal O_5$ generate nonvanishing cubic terms in the reduced action.
At the saddle point $\mu=\pi/2$, their contributions are
\begin{align}
\label{eq:G_i_mu_pi2}
\begin{aligned}
\frac{G_1(\pi/2,T)}{g v(T)}
&=
64\pi
\left(
\frac{v(T)}{\Lambda_{\rm dim8}}
\right)^4
\int_{0}^{\infty}d\xi\,
f(1-f)\frac{df}{d\xi}
\left[
3+h\left\{
6f^2h+(1-4f)(h+2)
\right\}
\right],
\\
\frac{G_2(\pi/2, T)}{g v(T)}
&=
128\pi
\left(
\frac{v(T)}{\Lambda_{\rm dim8}}
\right)^4
\int_{0}^{\infty}d\xi\,
f(1-f)^2
(2+h-3fh)
\frac{d}{d\xi}(hf),
\\
\frac{G_3(\pi/2,T)}{g v(T)}
&=
\frac{768\pi}{g}
\left(
\frac{v(T)}{\Lambda_{\rm dim8}}
\right)^4
\int_{0}^{\infty}d\xi\,
f^2(1-f)^2h^2
\frac{df}{d\xi},
\\
\frac{G_4(\pi/2,T)}{g v(T)}
&=
6144\pi
\left(
\frac{v(T)}{\Lambda_{\rm dim8}}
\right)^4
\int_{0}^{\infty}d\xi\,
\frac{f(1-f)}{\xi^2}
\frac{df}{d\xi}
\left[
\xi^2
\left(\frac{df}{d\xi}\right)^2
+
8f^2(1-f)^2
\right],
\\
\frac{G_5(\pi/2,T)}{g v(T)}
&=
2048\pi
\left(
\frac{v(T)}{\Lambda_{\rm dim8}}
\right)^4
\int_{0}^{\infty}d\xi\,
\frac{f(1-f)}{\xi^2}
\frac{df}{d\xi}
\left[
\xi^2
\left(\frac{df}{d\xi}\right)^2
+
8f^2(1-f)^2
\right],
\\
G_6(\mu, T)
&=
G_7(\mu, T)
=
0.
\end{aligned}
\end{align}
Thus, within the non-contractible-loop ansatz adopted here, $\mathcal O_6$ and $\mathcal O_7$ do not generate the cubic term responsible for the $CP$ asymmetry in the EW sphaleron process.

\subsection{Sphaleron-like configurations}

The ordinary EW sphaleron transition follows the minimum-energy path connecting neighboring vacua~\cite{Manton:1983nd,Akiba:1988ay}.
At finite temperature, however, thermal fluctuations can also explore deformed field configurations whose energies are higher than that of the sphaleron configuration.
Following Refs.~\cite{Hong:2023zrf}, we refer to transitions mediated by such configurations as sphaleron-like transitions.

We model the deformed configurations by rescaling the characteristic widths of the gauge and Higgs profiles as
\begin{align}
\label{eq:fh_ab}
\left.f(\xi)\right|_{\Xi\to\alpha\Xi_0} \,, \quad \left.h(\xi)\right|_{\Omega\to\beta\Omega_0} \,,
\end{align}
where $\alpha$ and $\beta$ parameterize deviations from the true sphaleron satisfying $\alpha = \beta = 1$.
For each deformed configuration, the corresponding actions are obtained through the replacements
\begin{align}
M(\mu, T) \to M(\mu,\alpha,\beta, T) \,, \quad 
G(\mu, T) \to G(\mu,\alpha,\beta, T) \,, \quad 
V(\mu, T) \to V(\mu,\alpha,\beta, T) \,.
\end{align}
Applying these replacements to Eq.~\eqref{eq:Acp_eff}, we obtain the
effective $CP$ asymmetry associated with each sphaleron-like transition as $A_{CP}^{\rm eff}(\alpha,\beta,T)$.

We then introduce a sphaleron-like transition rate,
$\Gamma_{\rm sph}(\alpha,\beta,T)$, normalized such that its integral over the deformation parameters reproduces the total sphaleron transition rate measured in lattice simulations~\cite{Hong:2023zrf, Ogawa:2026olw}
\begin{align}
\Gamma_{\mathrm{sph}}(\alpha, \beta, T)=\frac{e^{-V(\mu=\pi/2, \alpha, \beta, T) / T}}{\int d \alpha^{\prime} d \beta^{\prime} e^{-V\left(\mu=\pi/2, \alpha^{\prime}, \beta^{\prime}, T\right) / T}} \Gamma_{\mathrm{sph}}^{\mathrm{lattice}}(T) \,,
\end{align}
with 
\begin{align}
\label{eq:lattice_rate_normalization}
\Gamma_{\rm sph}^{\rm lattice}(T)
=
\iint d\alpha\,d\beta\,
\Gamma_{\rm sph}(\alpha,\beta,T) \,.
\end{align}
Here, $\Gamma_{\rm sph}^{\rm lattice}(T)$ denotes the sphaleron transition rate in the broken phase obtained in Refs.~\cite{DOnofrio:2014rug,Annala:2023jvr}.
In the following analysis, we employ the prescription developed in Ref.~\cite{Ogawa:2026olw}.


\section{Boltzmann equation for the baryon asymmetry}
\label{sec:BoltzmannEq}

As the temperature decreases during the EW crossover, sphaleron-like configurations with sufficiently small or large characteristic sizes gradually decouple from the thermal bath and subsequently decay~\cite{Kharzeev:2019rsy, Hong:2023zrf}.
The continuous decoupling of these configurations provides the departure from thermal equilibrium required by the third Sakharov condition~\cite{Sakharov:1967dj}.
The remaining Sakharov conditions are supplied by the EW sphaleron transitions, which violate baryon number through the chiral anomaly, and by the $CP$-violating operators, which generate an asymmetry between transitions toward neighboring topological vacua.
Consequently, the three Sakharov conditions can be satisfied simultaneously in the sphaleron decoupling in sphalerogenesis.

The evolution of the baryon number density $n_B$ is described by the following Boltzmann equation~\cite{Hong:2023zrf}
\begin{align}
\label{eq:BoltzmannEq}
-HT\frac{dn_B}{dT} + 3Hn_B = -\Gamma_B(T)n_B + P(T) \,,
\end{align}
where $\Gamma_B(T)$ and $P(T)$ denote the washout rate and the baryon-number source term, respectively.
We assume a radiation-dominated Universe where the Hubble parameter takes
\begin{align}
H(T) = \sqrt{\frac{\pi^2 g_*(T)}{90}} \frac{T^2}{M_{\rm Pl}} \,,
\end{align}
where $g_*(T)$ is the effective number of relativistic degrees of freedom and $M_{\rm Pl}$ is the reduced Planck mass.
The washout rate is given by~\cite{Tanaka:2025cpw,Ogawa:2026olw}
\begin{align}
\label{eq:GammaB}
\Gamma_B(T)
=
\begin{cases}
\displaystyle
\frac{39}{4T^3} \int_{S(T)} d\alpha \,d\beta\, \Gamma_{\rm sph}(\alpha,\beta,T),
& T_{\rm sph}<T<T_{\rm EW},
\\[2mm]
0 \,, 
& T<T_{\rm sph},
\end{cases}
\end{align}
while the source term is
\begin{align}
\label{eq:source}
P(T)
=
\begin{cases}
\displaystyle
\int_{\overline{S}(T)}
d\alpha\,d\beta\,
\Gamma_{\rm sph}(\alpha,\beta,T)\,
3A_{CP}^{\rm eff}(\alpha,\beta,T) \,,
&
T_{\rm sph}<T<T_{\rm EW} \,,
\\[2mm]
\displaystyle
\Gamma_{\rm sph}^{\rm lattice}(T)\,
3A_{CP}^{\rm eff}(1,1,T) \,,
&
T<T_{\rm sph} \,.
\end{cases}
\end{align}
Here, $S(T)$ denotes the region of the $(\alpha,\beta)$ plane in which the corresponding sphaleron-like transitions remain thermally active, whereas $\overline{S}(T)$ denotes the region containing configurations that have decoupled from the thermal bath (see Fig.~1 of Ref.~\cite{Ogawa:2026olw}).
The integral over $S(T)$ therefore determines the washout of a pre-existing baryon asymmetry, while the integral over $\overline{S}(T)$ describes baryon-number production associated with the decay of the decoupled sphaleron-like configurations.
The factor of $3$ accounts for the baryon-number change associated with a unit change in the Chern--Simons number in the SM with three fermion generations.
The detailed discussion relevant to the Boltzmann equation in Eq.~\eqref{eq:BoltzmannEq} is given in Ref.~\cite{Ogawa:2026olw}. 

The temperature $T_{\rm sph}$ denotes the conventional EW sphaleron freeze-out temperature obtained from lattice simulations~\cite{DOnofrio:2014rug,Annala:2023jvr}.
Below $T_{\rm sph}$, the washout process is neglected.
Following Refs.~\cite{Hong:2023zrf, Tanaka:2025cpw, Ogawa:2026olw}, the residual baryon-number source is approximated by the lattice-simulated sphaleron rate multiplied by the $CP$ asymmetry evaluated for the true sphaleron configuration with $\alpha=\beta=1$.
This contribution rapidly becomes negligible as the broken-phase sphaleron rate decreases at lower temperatures.

\begin{figure}[t]
    \centering
    \includegraphics[width=0.99\linewidth]{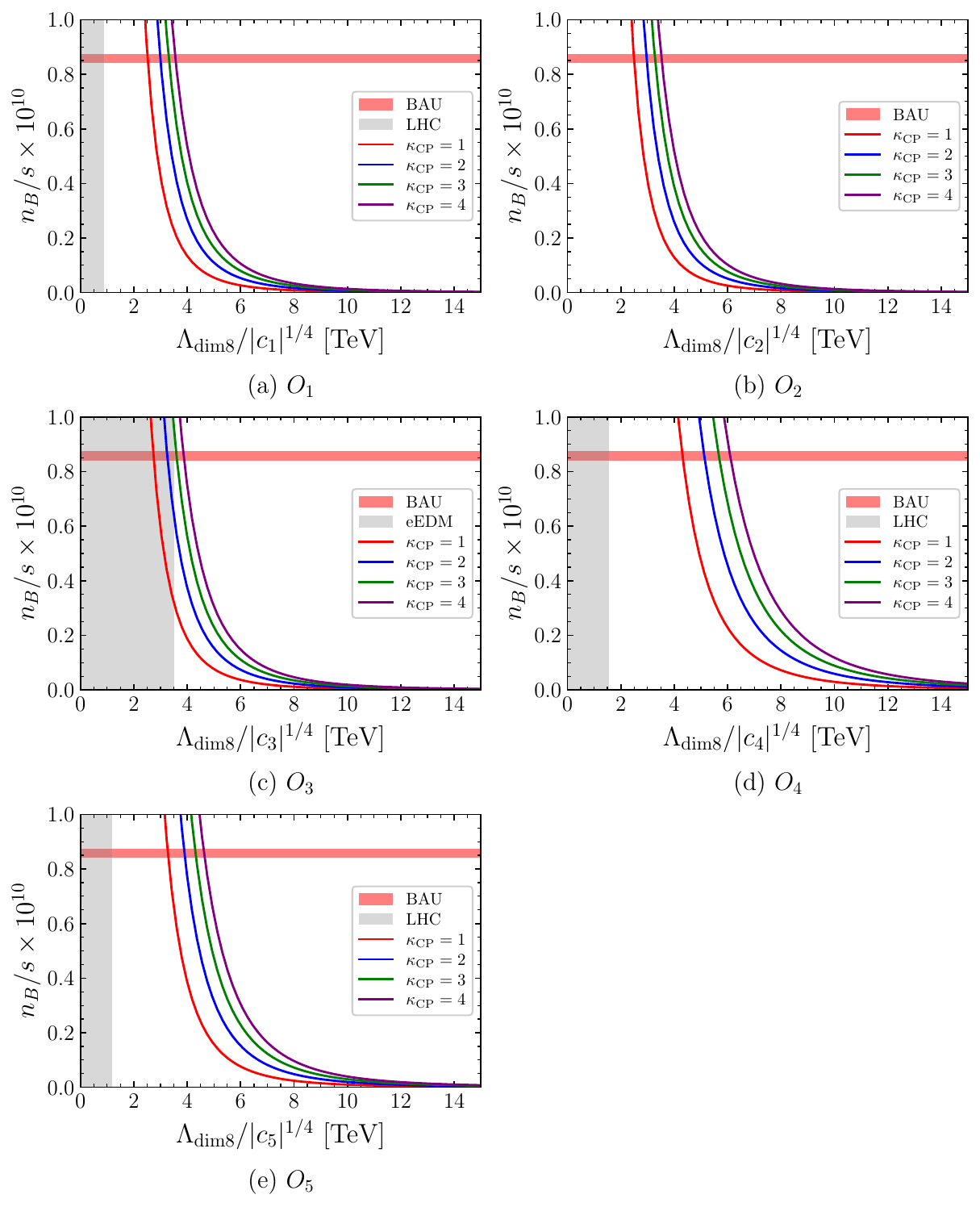}
    \caption{
    Baryon-to-entropy ratio generated by each dimension-8 operator as a function of the effective scale $\Lambda_{\rm dim8}/|c_i|^{1/4}$.
    The colored solid lines are the results with each $\kappa_{CP}$ value. 
    The red band denotes the observed baryon asymmetry given in Eq.~\eqref{eq:BAU}.
    The gray region indicates the experimental constraints by the LHC data~\cite{CMS:2026job} or the eEDM measurement~\cite{Roussy:2022cmp}.
    The positivity constraints discussed in the subsection~\ref{sec:positivity} are not imposed in this figure.
    }
    \label{fig:Cutoff_BAU}
\end{figure}

We numerically solve Eq.~\eqref{eq:BoltzmannEq} for each of the five dimension-8 operators that generate a nonvanishing $CP$ asymmetry.
Figure~\ref{fig:Cutoff_BAU} shows the resulting baryon-to-entropy ratio as a function of the effective cutoff scale $\Lambda_{\rm dim8}/|c_i|^{1/4}$.
The observed BAU can be reproduced if
\begin{align}
3\,{\rm TeV} \lesssim \frac{\Lambda_{\rm dim8}}{|c_i|^{1/4}} \lesssim 7\,{\rm TeV} \,,
\end{align}
with the precise range depending on the operator and on the benchmark value of $\kappa_{CP}$.
As shown in Fig.~\ref{fig:Cutoff_BAU}, the variation among the five operators is relatively modest.
This can be understood from the scaling $A_{CP}^{\rm eff} \propto c_i/\Lambda_{\rm dim8}^4.$
Differences in the numerical coefficients and radial integrals entering $G_i$ can therefore be compensated by comparatively small shifts in the effective scale $\Lambda_{\rm dim8}/|c_i|^{1/4}$.

For the case with $\mathcal{O}_{1}, \mathcal{O}_{4}$ and $\mathcal{O}_{5}$, the gray region is constrained by the measurement of the photon-fusion production of $W$ boson pairs using the data at the LHC with $138\,\text{fb}^{-1}$ and $\sqrt{s} =13\,{\rm TeV}$~\cite{CMS:2026job}. 
No direct collider limit is imposed on $\mathcal{O}_{2}$ in Fig.~\ref{fig:Cutoff_BAU} because the photon-fusion channel analyzed in Ref.~\cite{CMS:2026job} does not provide a bound on this operator direction.
For the operator $\mathcal{O}_{3}$, it is strongly constrained by the electron electric dipole moment (eEDM). 
At the zero temperature, the operator $\mathcal{O}_{3}$ takes the same expression of the EW-Weinberg operator in Eq.~\eqref{eq:Scp6} with the replacement
\begin{align}
\label{eq:Ldim6_Ldim8}
\frac{1}{\Lambda_{\rm dim6}^2} \to \frac{3v^2 c_{3}}{2g \Lambda_{\rm dim8}^4} \,. 
\end{align}
On the other hand, the EW-Weinberg operator induces the deviation in the eEDM as~\cite{Boudjema:1990dv}
\begin{align}
\label{eq:de_dim6}
\left|\frac{d_{e}}{e}\right| = \frac{g^2 m_{e}}{96 \pi^2 \Lambda_{\rm dim6}^2} \simeq 4.1 \times 10^{-30}\,{\rm cm} \left( \frac{33\,{\rm TeV}}{\Lambda_{\rm dim6}} \right)^2 \,. 
\end{align}
Performing the replacement in Eq.~\eqref{eq:Ldim6_Ldim8}, the eEDM induced by the operator $\mathcal{O}_{3}$ is given by
\begin{align}
\label{eq:de_O3}
\left| \frac{d_{e}}{e} \right| = \frac{3\, g\, m_{e} |c_{3}| v^2}{192 \pi^2 \Lambda_{\rm dim8}^4} \simeq 4.1 \times 10^{-30}\,{\rm cm} \left( \frac{3.5\,{\rm TeV}}{\Lambda_{\rm dim8}/|c_{3}|^{1/4}} \right)^4 \,. 
\end{align}
In our analysis, we regard Eq.~\eqref{eq:de_O3} as the leading contribution to the eEDM from $\mathcal{O}_{3}$ and take into account the current strongest experimental constraint $|d_{e}/e| < 4.1 \times 10^{-30}\,{\rm cm}$ reported by JILA~\cite{Roussy:2022cmp}.


\section{Loop counting and implications for UV matching}
\label{sec:UVcompletion}

As emphasized in Refs.~\cite{Arzt:1994gp,Buchalla:2022vjp}, a consistent estimate of new physics effects in the SMEFT requires not only canonical mass-dimension counting but also loop-order counting.
This observation is particularly relevant to sphalerogenesis because the dimension-8 operators discussed above may modify the baryon asymmetry predicted from the EW-Weinberg operator studied in Refs.~\cite{Tanaka:2025cpw,Ogawa:2026olw}.
In this section, we estimate the relative importance of these contributions from the viewpoint of UV matching.

In renormalizable UV completions, the EW-Weinberg operator first arises at two-loop order~\cite{Banno:2024apv,Banno:2026hsc}.
Neglecting model-dependent couplings and numerical matching factors, its effective cutoff scale can therefore be estimated as
\begin{align}
\label{eq:Lambda_dim6}
\Lambda_{\rm dim6} \sim (4\pi)^2 M_{\rm new} \,,
\end{align}
where $M_{\rm new}$ denotes the characteristic physical mass scale of the new particles propagating in the loops.
If the same UV theory also generates one of the dimension-8 operators in Eq.~\eqref{eq:dim8_OPEs} at $n$-loop order, we parametrize its cutoff scale as
\begin{align}
\label{eq:Lambda_dim8}
\Lambda_{\rm dim8} \sim (4\pi)^{n/2}\zeta M_{\rm new} \,,
\end{align}
where $\zeta$ is a dimensionless model-dependent matching factor.

To compare the two contributions, we define an operator-dependent numerical coefficient $k_i$ through
\begin{align}
\left.
\frac{G_i}{G_{\rm dim6}}
\right|_{\mu=\pi/2}
=
k_{i}
\frac{\Lambda_{\rm dim6}^2 v^2(T)}
{\Lambda_{\rm dim8}^4}.
\end{align}
The coefficient $k_i$ is determined by the numerical prefactors and integrals of the radial profile entering the reduced sphaleron action.
For the five operators considered here, it is typically in the range $k_i=\mathcal{O}(1\text{--}100)$.
Using Eqs.~\eqref{eq:Lambda_dim6} and \eqref{eq:Lambda_dim8}, the ratio becomes
\begin{align}
\label{eq:G_ratio_loop}
\left.
\frac{G_i}{G_{\rm dim6}}
\right|_{\mu=\pi/2}
=
\frac{k_i}{\zeta^2} (16\pi^2)^{1-n} \left( \frac{4\pi v(T)/\zeta}{M_{\rm new}} \right)^2.
\end{align}

\begin{figure}[t]
    \centering
    \includegraphics[width=0.97\linewidth]{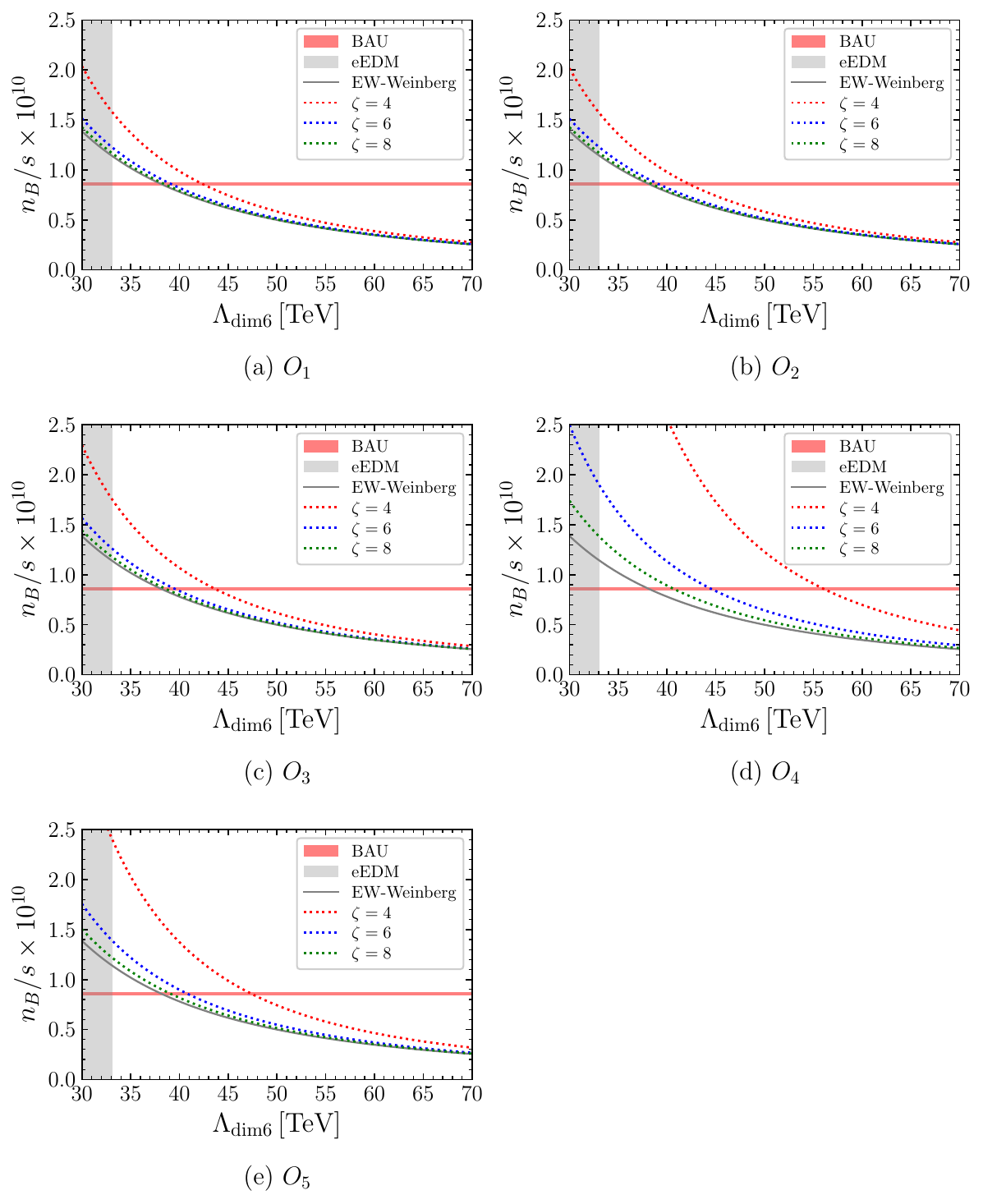}
    \caption{
    Baryon-to-entropy ratio as a function of $\Lambda_{\rm dim6}$ in the presence of the EW-Weinberg operator and one dimension-8 operator.
    We set $c_i=1$, $n=1$, and $\kappa_{CP}=1$.
    The colored dotted curves correspond to different values of the matching factor $\zeta$, while the gray solid curve includes only the EW-Weinberg contribution.
    The gray region shows the eEDM constraint based on Eq.~\eqref{eq:de_dim6} without dimension-8 operator contributions. 
    The red band is the observed baryon asymmetry given in Eq.~\eqref{eq:BAU}.
    The positivity constraints discussed in the subsection~\ref{sec:positivity} are not imposed.
    }
    \label{fig:EW_Weinberg_w_dim8}
\end{figure}

Equation~\eqref{eq:G_ratio_loop} explicitly shows the competition between canonical-dimension suppression and loop enhancement.
In particular, it indicates that the dimension-8 operator contributions to sphalerogenesis can be comparable to that of the EW-Weinberg operator if $n = 1$ and $M_{\rm new} \sim 4 \pi v/\zeta$ with $\zeta \sim \mathcal{O}(1-10)$. 
This finding indicates that $CP$-violating dimension-8 operators may be more significant than the naive expectation based on the canonical dimension counting. 
Indeed, the UV realizations studied in Ref.~\cite{Ogawa:2026olw} often favor
$M_{\rm new}\lesssim\mathcal O(1)~{\rm TeV}$.
Consequently, one-loop-generated $CP$-violating dimension-8 operators can provide contributions comparable to, or even larger than, the two-loop-induced EW-Weinberg contribution.
Therefore, the canonical mass-dimension counting alone may therefore underestimate their importance in sphalerogenesis.

Figure~\ref{fig:EW_Weinberg_w_dim8} shows the predicted baryon-to-entropy ratio as a function of $\Lambda_{\rm dim6}$ when the EW-Weinberg operator is supplemented by each of the five dimension-8 operators.
The dimension-8 cutoff scale is related to the new particle mass through Eq.~\eqref{eq:Lambda_dim8}.
We take $c_i=1$, $n=1$, and $\kappa_{CP}=1$ as representative benchmark choices and consider several values of $\zeta$.
The gray solid curve shows the result obtained from the EW-Weinberg operator alone, while the colored dotted curves include the additional dimension-8 contribution.
The gray region shows the constraint from the eEDM measurement with Eq.~\eqref{eq:de_dim6} without the contribution from dimension-8 operators. 
The dimension-8 operators can substantially shift the value of $\Lambda_{\rm dim6}$ required to reproduce the observed baryon asymmetry.
The direction and magnitude of this shift generally depend on the relative signs and matching coefficients of the dimension-6 and dimension-8 operators.
The benchmark shown in Fig.~\ref{fig:EW_Weinberg_w_dim8} corresponds to constructive interference. 

Figure~\ref{fig:EW_Weinberg_w_dim8} demonstrates that dimension-8 contributions cannot generally be neglected when making quantitative predictions for sphalerogenesis in a specific UV theory.
Although this conclusion is similar to the results of EW baryogenesis in the SMEFT framework~\cite{Bodeker:2004ws,Huber:2006ri,Huang:2015izx,deVries:2017ncy}, its interpretation is different.
In conventional EW baryogenesis, a relatively low cutoff scale is often required to realize a sufficiently strong first-order EW phase transition, so that higher-dimensional operators can become important for maintaining the consistency of the EFT expansion.
On the other hand, in sphalerogenesis, dimension-8 operators can compete with the EW-Weinberg operator if the two contributions may arise at different loop orders, even when the canonical mass-dimension expansion alone would suggest a stronger suppression of the dimension-8 operators.

\begin{figure}[t]
    \centering
    \includegraphics[width=0.97\linewidth]{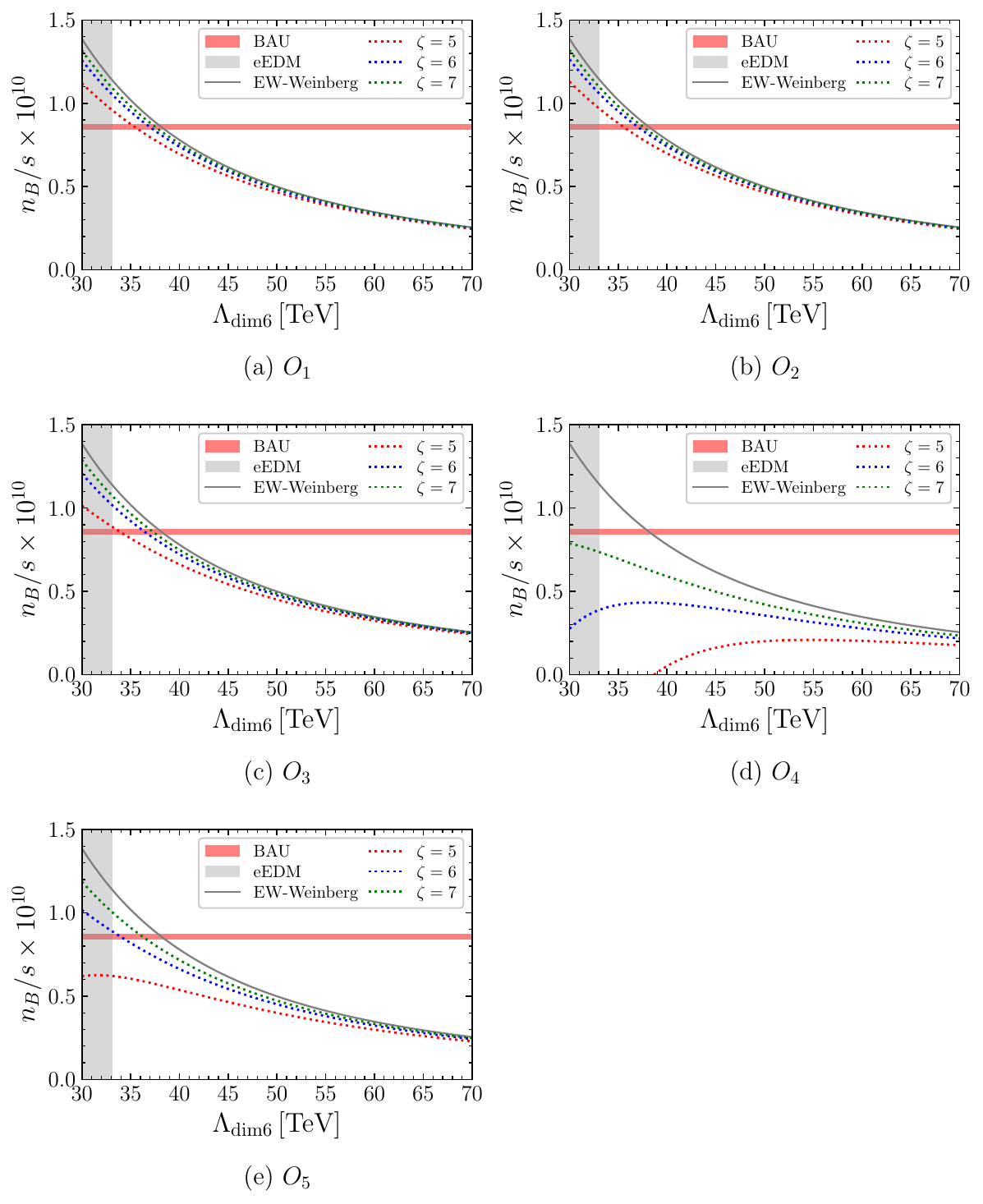}
    \caption{
    Baryon-to-entropy ratio as a function of $\Lambda_{\rm dim6}$ in the presence of the EW-Weinberg operator and one dimension-8 operator.
    The definition of colored lines and regions is the same as Fig.~\ref{fig:EW_Weinberg_w_dim8}.
    We set $c_i=-1$, $n=1$, and $\kappa_{CP}=1$.
    The relative sign is chosen such that the dimension-6 and dimension-8 contributions interfere destructively.
    The positivity constraints discussed in the subsection~\ref{sec:positivity} are not imposed.
    }
    \label{fig:EW_Weinberg_w_dim8_cm1}
\end{figure}

In Figure~\ref{fig:EW_Weinberg_w_dim8_cm1}, it shows the predicted baryon-to-entropy ratio as a function of $\Lambda_{\rm dim6}$ with dimension-8 operators that cause destructive interference. 
In this plot, we take $c_i=-1$, $n=1$, and $\kappa_{CP}=1$ and consider several values of $\zeta$.
The definition of colored lines and regions is the same as Fig.~\ref{fig:EW_Weinberg_w_dim8}.
In particular, for the operators $\mathcal{O}_{4}$ and $\mathcal{O}_{5}$, the BAU may not be feasible on any cutoff scale $\Lambda_{\rm dim6}$ when the value of $\zeta$ is small. 
Although these results could potentially be improved by considering operators of even higher-mass dimension, that is beyond the scope of this study.

We emphasize that the analysis presented here is a power-counting estimate.
For UV mass scales only moderately above the EW temperature, the validity of the EFT expansion and the complete set of correlated matching contributions should be examined within an explicit UV model.


\section{Discussion}
\label{sec:discussions}

We comment on the experimental and theoretical constraints relevant to our scenario.

\subsection{Electron electric dipole moment constraints} \label{sec:eEDM}

In this work, we have considered the eEDM constraint only for the operator $O_3$. 
The remaining operators are also expected to be constrained by eEDM measurements, although a quantitative analysis is beyond the scope of this work. 
Indeed, in several UV-complete theories, operators that do not contribute to sphalerogenesis can nevertheless be generated together with operators relevant for sphalerogenesis and may induce additional contributions to the eEDM. 
Therefore, a comprehensive analysis requires matching a specific UV completion onto the complete set of effective operators and evaluating all EDM contributions consistently. 
We leave such an analysis to future work.

\subsection{Positivity bounds} \label{sec:positivity}

Dimension-8 operators generated by a consistent UV completion are subject to theoretical constraints arising from causality, unitarity, and the analytic structure of scattering amplitudes~\cite{Adams:2006sv,Remmen:2019cyz,Yamashita:2020gtt}.
The same constraint can also be obtained by considering the non-negativity of relative entropy~\cite{Cao:2022iqh}.
Within the operator sector and forward-limit analysis considered in Ref.~\cite{Remmen:2019cyz}, the $CP$-odd operators $\mathcal O_4$ and $\mathcal O_5$ cannot be introduced independently when the corresponding $CP$-even operators are absent.
In particular, their Wilson coefficients must vanish if only one of the two operators is present.

When $\mathcal O_4$ and $\mathcal O_5$ are included simultaneously, a nonvanishing $CP$-odd direction is allowed if their Wilson coefficients satisfy
\begin{align}
\label{eq:positivity_c4c5}
c_4=-c_5.
\end{align}
The corresponding operator combination is therefore $\mathcal{O}_4-\mathcal{O}_5 = \mathcal{O}_{4-5}$. 
In contrast, the forward-limit positivity analysis of Ref.~\cite{Remmen:2019cyz} does not yield analogous nontrivial constraints on $\mathcal O_1$, $\mathcal O_2$, and $\mathcal O_3$.
Consequently, the four operators $\mathcal O_1, \mathcal{O}_2, \mathcal{O}_3$ and $\mathcal{O}_{4-5}$ can provide independent sources of the $CP$ asymmetry in the EW sphaleron process in our scenario.

For the cubic term in the reduced sphaleron action in Eq.~\eqref{eq:Ssph_dim8}, the positivity-compatible combination $\mathcal{O}_{4-5}$ gives
\begin{align}
\label{eq:G45}
G_{4-5}(\pi/2) \equiv G_4(\pi/2)-G_5(\pi/2) = \frac{2}{3}G_4(\pi/2) \,, 
\end{align}
where we have used $G_4(\pi/2)=3\,G_5(\pi/2)$, as follows from Eq.~\eqref{eq:G_i_mu_pi2}.
Thus, the positivity-compatible operator combination remains a nonvanishing source of the $CP$ asymmetry and can generate the baryon asymmetry through sphalerogenesis.


\section{Conclusions} \label{sec:conclusions}

We have investigated sphalerogenesis within the SMEFT including the $CP$-violating dimension-8 operators listed in Eq.~\eqref{eq:dim8_OPEs}.
Among the seven operators constructed from the Higgs doublet and the $SU(2)_L$ gauge fields, five generate a nonvanishing $CP$-odd cubic term in the reduced sphaleron action and can therefore act as sources of the $CP$ asymmetry in EW sphaleron-like transitions.

By solving the Boltzmann equation for the baryon asymmetry, we have shown that the observed BAU can be reproduced for effective cutoff scales in the approximate range $3\,{\rm TeV} \lesssim \Lambda_{\rm dim8}/|c_i|^{1/4} \lesssim 7\,{\rm TeV}$, where the precise value depends on the operator and on the uncertainty associated with the reduced sphaleron description.

We have also examined the dimension-8 contributions in the presence of the dimension-6 operator.
Since the EW-Weinberg operator is generated at two-loop order in renormalizable UV completions~\cite{Banno:2024apv, Banno:2026hsc}, dimension-8 operators generated at one loop can provide a contribution comparable to, or even larger than, the dimension-6 contribution despite its higher canonical dimension.
This demonstrates that loop-order counting, in addition to canonical mass-dimension counting, is essential for quantitatively assessing sphalerogenesis within the SMEFT.

Our results therefore show that $CP$-violating dimension-8 operators can play an important role both as independent sources of the baryon asymmetry and as corrections to predictions based on the EW-Weinberg operator.
A complete assessment in a specific UV completion requires the correlated matching of the dimension-6 and dimension-8 operator coefficients, together with dedicated experimental constraints on the relevant $CP$-violating interactions such as eEDM measurements.

\section*{Acknowledgment}
K.O. is supported by the ``Make New Standards Program for the Next Generation Researchers'' of the Tokai National Higher Education and Research System (THERS).
This work was also financially supported by JST SPRING, Grant Number JPMJSP2125.

\bibliographystyle{JHEP}
\bibliography{reference} 


\end{document}